\documentclass[twocolumn, showpacs, superscriptaddress, a4paper]{revtex4}%
\usepackage{amsfonts}
\usepackage{amsmath}
\usepackage{amssymb}
\usepackage{dcolumn}
\usepackage{bm}
\usepackage{graphicx}
\usepackage{geometry}%
\providecommand{\U}[1]{\protect\rule{.1in}{.1in}}
\begin{document}
\title{Quantum walks in commensurate off-diagonal Aubry-Andr\'{e}-Harper model}
\author{Li Wang}
\email{liwangiphy@sxu.edu.cn}
\affiliation{Institute of Theoretical Physics, Shanxi University, Taiyuan 030006, P. R. China}
\affiliation{Department of Theoretical Physics, Research School of Physics and Engineering, Australian National University, Canberra, ACT 0200, Australia}
\author{Na Liu}
\affiliation{Institute of Theoretical Physics, Shanxi University, Taiyuan 030006, P. R. China}

\author{Shu Chen}
\affiliation{Beijing National
Laboratory for Condensed Matter Physics, Institute of Physics,
Chinese Academy of Sciences, Beijing 100190, China}
\affiliation{School of Physical Sciences, University of Chinese Academy of Sciences, Beijing, 100049, China}
\affiliation{Collaborative Innovation Center of Quantum Matter, Beijing, China}
\author{Yunbo Zhang}
\affiliation{Institute of Theoretical Physics, Shanxi University, Taiyuan 030006, P. R. China}

\date{\today}

\begin{abstract}

Due to the topological nature of Aubry-Andr\'{e}-Harper (AAH) model, exotic edge states have been found existing in one-dimensional periodic and quasiperiodic lattices. In this article, we  investigate continuous-time quantum walks of identical particles initially located on either edge of commensurate AAH lattices in detail. It is shown that the quantum walker is delocalized
among the whole lattice until the strength of  periodic modulation is strong enough. The inverse participation ratios (IPRs) for all of the eigenstates  are calculated. It is found that the localization properties of the quantum walker is mainly determined by the IPRs of the topologically protected edge states. More interestingly, the edge states are shown to have an exotic `\emph{repulsion}' effect on quantum walkers initiated from the lattice sites inside the bulk.  Furthermore, we examine the role of nearest-neighbour interaction on the quantum walks of two identical fermions. Clear enhancement of the `\emph{repulsion}' effect by strong interaction has been shown.

\end{abstract}

\pacs{37.10.Jk, 05.60.Gg, 05.40.Fb}



\maketitle

\bigskip

\section{INTRODUCTION}

Quantum walks,  the quantum analog of classical random walks,  describe the random dynamics of quantum particles on a discrete lattice\cite{Aharonov,kempe}, which is inherently governed by the time-dependent wavefunction of the system. Compared to the classical random walks, dramatically different behavior shows in quantum walks due to the coherent superposition and interference of the wavefunction. For example, it is well-known now that a quantum walker can propagate linearly with respect to the expansion time, which is much faster than its classical counterpart. This may be exploited in the designing of more efficient quantum search algorithm for quantum compuation\cite{kempe, Ambainis,Childs2009, Childs2013, gutmann}.
Having witnessed the huge success of  classical random walks, people believe that quantum walks may have widespread applications in quantum algorithms\cite{kempe,Ambainis}, quantum computing\cite{Childs2013}, quantum information\cite{Venegas}, quantum simulation\cite{Asboth}, quantum biology\cite{Lloyd} and so on. Motivated by this promising prospect, more and more research activities on quantum walks have been undertaken  by both experimentalists and theorists. Actually, quantum walks have been experimentally implemented in a variety of quantum systems\cite{jbbook}, such as optical resonator\cite{Bouwmeester}, nuclear magnetic resonance\cite{Duj}, trapped ions\cite{Schmitz}, trapped cold neutral atoms\cite{Karski,greiner15}, single photons in bulk\cite{Broome}, fiber optics\cite{Schreiber}, and coupled waveguide arrays\cite{Sansoni}. And on the theoretical side, fundamental effects of quantum statistics\cite{Bordone, qinxz}, interactions\cite{qinxz, Lahini, wangepjd, liwang}, disorders\cite{Yiny,BordoneA}, defects\cite{zjpra,zjsci},  and hopping modulations\cite{zjsci,liwang,liwang15,Kraus}  on quantum walks have been intensively investigated.

In this article, we investigate the quantum walks of one and two identical fermions on a one-dimensional optical lattice with periodically modulated hoppings, which is described by the commensurate off-diagonal Aubry-Andr\'{e}-Harper (AAH) model.
The original version of AAH model\cite{AA,Harper} which contains an incommensurate potential was initially introduced to  study the localization phenomena in one-dimension. Compared to the one-dimensional Anderson localization model\cite{Anderson}, AAH model features the appearance of a non-trivial localization transition which is more interesting from the perspective of physics of disorder-induced metal-insulator transition.
Afterwards, the original AAH model has been generalized to a variety of versions adapting  to different physical problems. For instance, AAH model with pure off-diagonal couplings has been used to investigate topological adiabatic pumping\cite{Kraus, liufangli,leechaohong}. And it has been shown that AAH model with on-site and/or off-diagonal modulation is topologically equivalent to Fibonacci lattices of the same quasiperiodicity\cite{KrausE, Verbin}. Topologically protected edge states have been found both in incommensurate\cite{Kraus} and commensurate\cite{Lang, dasama,guo} AAH models.

Here in this work, we first look at the quantum walks of identical particles initially located at either boundary of a period-2 commensurate off-diagonal AAH lattice with topologically nontrivial edge states 
 to investigate the effect of periodic hopping modulations. It is found that the quantum walker is delocalized among the whole lattice until the strength of the modulation on the hopping term is strong enough. However, for topologically trivial phase with the phase factor $\phi$ of the hopping modulations varying outside the interval $(-\pi/2,\pi/2)$, the quantum walker is always delocalized. 
 It is shown that this phenomena is attributed to the topological properties of the commensurate off-diagonal  AAH model. We calculate the inverse participation rations (IPRs) for all of the eigenstates and find that the localization of the quantum walker is mainly determined by the IPRs of the exotic edge states. Secondly, we examine the quantum walks of identical particles initially setting out from lattice sites in the bulk. It turns out that quantum walker initially located on any lattice site in the bulk may expand ballistically as usual and no evidence of localization shows as the strength of the hopping modulation varies. However, an interesting and subtle phenomenon emerges when we look at either boundary of the lattice. It seems that the topologically protected edge state has an exotic \emph{repulsion} effect which make the lattice boundary unreachable for quantum walker setting out from bulk sites. Thirdly, we investigate quantum walks of two identical fermions with  nearest-neighbor interaction and find that strong interactions enhance the exotic \emph{repulsion} effect of the edge states. Brief discussions on the experimental realization of these effects and their potential applications in the future's quantum information techniques are given therein.

The paper is organized as follows. In Sec. II, we introduce the commensurate off-diagonal AAH model. Nearest-neighbor interaction is also considered. We construct the
Hilbert space for quantum walkers and briefly show the method we use to describe the time-evolution of the density distribution of quantum walkers.  Single-particle quantum walks in commensurate off-diagonal AAH model are shown in Sec. III.  Time-evolution of the density distributions and IPRs of all the eigenstates are calculated. Detailed  analysis corresponding to the dynamical properties is addressed therein.
In Sec. IV, we turn to investigate quantum walks of two identical fermions with nearest-neighbor interactions.  Finally, a brief summary is given in Sec. V.

\section{Model and Method\label{secii}}

We investigate the continuous-time quantum walks of one and two identical fermions on a one-dimensional lattice with periodically modulated hoppings. The dynamics of such a system is governed by the so-called commensurate off-diagonal AAH model. Additionally, nearest-neighbor interaction between particles is also considered. Therefore, the Hamiltonian of this system reads,

\begin{align}
H=\sum_{i} J_i    \hat{c}^{\dagger}_{i+1} \hat{c}_{i}+V \sum_{i}  \hat{n}_{i} \hat{n}_{i+1},
\label{H}%
\end{align}
with
\begin{align}
J_i=t+\lambda_{od} \cos(2\pi i / T+\phi),
\label{Ji}%
\end{align}
where $\hat{c}_i^{\dagger}$ ($\hat{c}_i$) is the creation (annihilation) operator of fermions, and $\hat{n}_i=\hat{c}_i^{\dagger}\hat{c}_i$ denotes the particle number of ferimons on site $i$. The hopping amplitude $t$ is set to be the unit of energy ($t=1$). All other parameters are scaled by $t$ in the following numerical and analytical investigations. $\lambda_{od}$ describes the strength of the cosine off-diagonal modulation, $T$ is the periodicity of the modulation, $\phi$ is a phase factor and $V$ is the strength of the nearest-neighbor interaction between particles. Since in this paper the AAH lattice we considered is commensurate, $T$ is set to be an integer. Specifically, our analysis and discussions in the following are mainly based on the period-2 case, i.e., $T=2$. And the lattices we studied in this paper are all finite.


To investigate the dynamics of quantum walkers initially located on well defined sites of the commensurate off-diagonal one-dimensional AAH lattice of length $L$, we resort to numerical techniques to solve the time-dependent Schr\"{o}dinger equation exactly. Since
$\left[  N,H\right]  =0$, the total particle number $N=\sum_{i}\hat{n}_i$ is conserved and the
system will evolve in the Hilbert space with fixed particle number. For single-particle quantum walks, the Hilbert space involved is simply spanned by basis $B^{(1)}=\{ \left | i \right >=c_{i}^{\dagger} \left | \mathbf{0} \right >, 1\leqslant i \leqslant L\}$, where $ \left | \mathbf{0} \right >$ denotes the vacuum state. With these basis, it is easy to construct the single-particle Hamiltonian matrix $H^{(1)}$. In units of $\hbar=1$,
the time evolution of an arbitrary single-particle state $\left\vert \psi^{(1)} \left(  t\right)  \right\rangle$ obeys time-dependent Schr\"{o}dinger equation,
\begin{equation}
i\frac{d}{dt}\left\vert \psi^{(1)} \left(  t\right)  \right\rangle =H^{\left(
1\right)  }\left\vert \psi^{(1)} \left(  t\right)  \right\rangle ,
\label{hs}
\end{equation}
with $\left\vert \psi^{(1)} \left(  t \right)  \right\rangle = \sum_{i}  a_{i}(t)  \left\vert i \right\rangle $. Similarly, for quantum walks of two identical fermions, the Hilbert space is spanned by the basis
$B^{(2)}=\{ \left | ij \right >=c_{i}^{\dagger}c_{j}^{\dagger} \left | \mathbf{0} \right >, 1\leqslant i < j \leqslant L \}$. And the the time evolution of an arbitrary two-particle state $\left\vert \psi^{(2)} \left(  t\right)  \right\rangle$ obeys,
\begin{equation}
i\frac{d}{dt}\left\vert \psi^{(2)} \left(  t\right)  \right\rangle =H^{\left(
2\right)  }\left\vert \psi^{(2)} \left(  t\right)  \right\rangle,
\label{H2}
\end{equation}
where $\left\vert \psi^{(2)} \left(  t \right)  \right\rangle = \sum_{i<j}  a_{ij} (t)  \left\vert i j \right\rangle $.

By solving the time-dependent equation (\ref{hs}) or (\ref{H2}) numerically, the wavefunction $\left\vert \psi^{1} \left(  t\right)  \right\rangle$ or $\left\vert \psi^{2} \left(  t\right)  \right\rangle$ which governs the dynamics of quantum walkers on the AAH lattice is obtained. Therefore, the time-dependent density distribution of quantum walkers is given by
\begin{align}
\left < n_{i}^{(s)}(t) \right> = \left< \psi^{(s)} (t) \left | \hat{c}_{i}^{\dagger}  \hat{c}_{i} \right |  \psi^{(s)} (t) \right>,
\end{align}
with $s=1$ or $2$ corresponding to  single-particle or two-particle quantum walk.


\begin{figure}[t]
\includegraphics[width=0.5\textwidth]{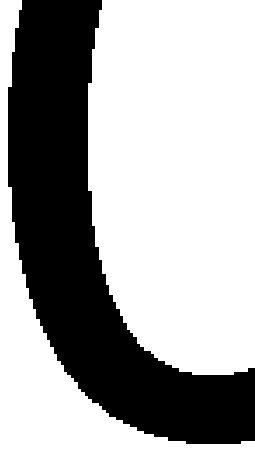} \caption{(Color online)
Single-particle quantum walks on a one-dimensional period-2 off-diagonal AAH lattice with $L=100$.
The quantum walker is initially positioned on the left boundary site.
(a-c) $\phi=0$, open boundary condition.
(d-f) $\phi=0.6\pi$, open boundary condition.
(g-i) $\phi=0$, periodic boundary condition.
The first column is corresponding to $\lambda_{od}=0.1$, the second is  $\lambda_{od}=0.3$ and the third is $\lambda_{od}=0.9$.
}
\label{fig1}
\end{figure}

\section{single-particle quantum walks} \label{sfqw}

Firstly, we investigate continuous-time quantum walks of single particles initially located on either boundary of an off-diagonal AAH lattice with $T=2$. The corresponding results are shown in Fig.\ref{fig1}. The length of the one-dimensional AAH lattice is $L=100$.
Fig.\ref{fig1}(a-f) is for AAH lattice with open boundary condition, while Fig.\ref{fig1}(g-i) is under periodic boundary condition.
In Fig.\ref{fig1}(a-c) and (g-i), the value of the off-diagonal modulation phase is $\phi=0$, and in Fig.\ref{fig1}(d-f), the phase $\phi$ is chosen to be $0.6\pi$.

\begin{figure}[t]
\includegraphics[width=0.48\textwidth]{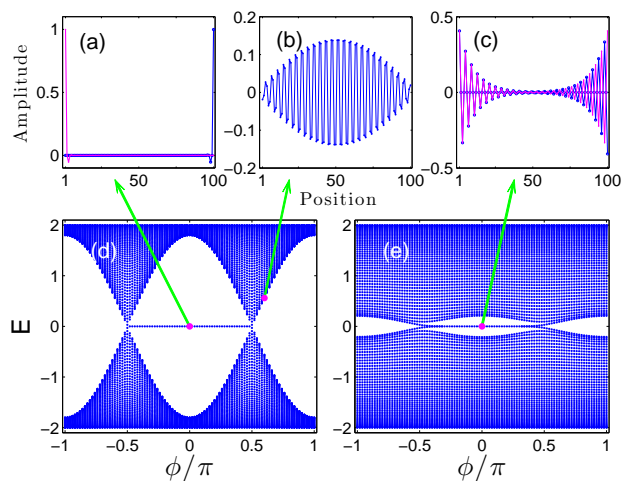} \caption{(Color online)
Upper panel shows the edge states of the one-dimensional period-2 off-diagonal AAH lattice with $L=100$ under open boundary condition. The lower panel is the energy spectrum plotted as a function of $\phi$.
(a,b,d) $\lambda_{od}=0.9$.
(c,e) $\lambda_{od}=0.1$.}
\label{fig2}
\end{figure}

It is found that the quantum walker is well localized on boundary site of the AAH lattice for sufficiently strong off-diagonal modulation, see Fig.\ref{fig1}(c) with $\lambda_{od}=0.9$. In order to observe the localization phenomenon more clearly, only $30$ sites are shown. This interesting localization phenomenon\cite{Kraus} is attributed to the appearance of exotic edge states \cite{dasama} in the energy spectrum of AAH lattices. As shown in Fig.\ref{fig2}(d), a pair of edge states indeed appear in the energy spectrum of the AAH model with $\lambda_{od}=0.9$. The probability amplitude distribution of corresponding eigenstates are shown in Fig.\ref{fig2}(a).

\begin{figure}[b]
\includegraphics[width=0.36\textwidth]{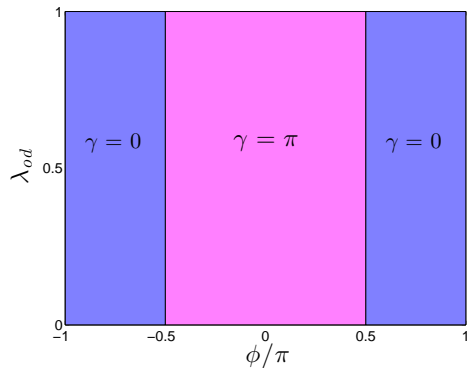} \caption{(Color online)
Phase diagram of the one-dimensional period-2 off-diagonal AAH model.
$\gamma$ is the Zak phase plotted as a function of the phase $\phi$ and the strength $\lambda_{od}$ of the off-diagonal modulations.
}
\label{fig3}
\end{figure}

\begin{figure}[t]
\includegraphics[width=0.52\textwidth]{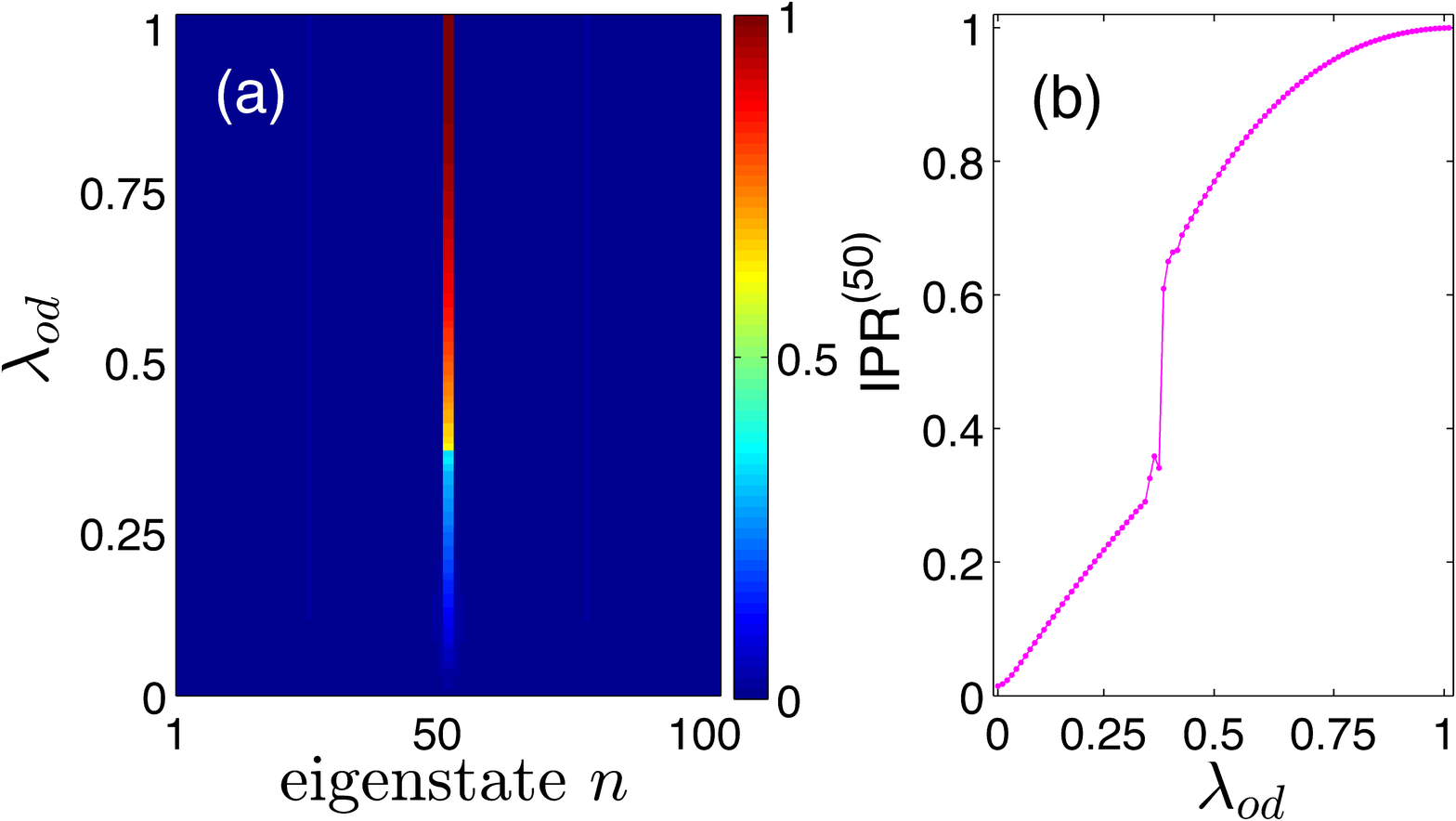} \caption{(Color online)
(a) Inverse participation ratio of all eigenstates plotted as a function of the strength of the off-diagonal modulations $\lambda_{od}$ for the one-dimensional period-2 off-diagonal AAH model  under open boundary condition with $\phi=0$ and $L=100$.
(b) Inverse participation ratio of the edge state as a function of $\lambda_{od}$.
}
\label{fig4}
\end{figure}

As we have seen in Fig.\ref{fig2}(d), the interesting edge states only occur in the regime with $-\pi/2<\phi<\pi/2$. This is actually determined by the topological properties of the commensurate off-diagonal AAH model, which could be characterized by Zak phase\cite{zak, niuq, linhu} i.e. the one-dimensional Berry phase across the Brillouin zone. The Zak phase is explicitly defined as
\begin{align}
\gamma=i \int_{BZ} dk \left< \Phi (k) \left |  \frac{d}{dk} \right |  \Phi(k) \right>,
\end{align}
where $\Phi(k)$ is the eigenstate of the occupied Bloch band.
In Fig.\ref{fig3}, we have calculated the Zak phase of the commensurate off-diagonal AAH model. It is shown that this model is topologically nontrivial in the regime $\phi \in (-\pi/2,\pi/2)$ characterized by $\gamma=\pi$. And it turns out that the Zak phase is insensitive to the strength of the off-diagonal modulations. Therefore, for weak off-diagonal modulations, edge states also appear in the energy spectrum under open boundary condition according to the bulk-edge correspondence, as is shown in Fig.\ref{fig2}(e).

However, the quantum walker is not well localized until the off-diagonal modulation is strong enough, see Fig.\ref{fig1}(a-c).
To quantify the localization property of the quantum walker, we compute the inverse participation ratio (IPR) for all of the eigenstates of off-diagonal AAH model with $T=2$ and $\phi=0$. For an eigenstate $\varphi_n$, which is spanned as $\varphi_n =\sum_{i} u^{n}_{i} \left |  i \right>$ in the single-particle Hilbert space $B^{(1)}$, the IPR\cite{kramer} is defined as
\begin{align}
I\!P\!R^{(n)}=\frac{\sum_{i} \left | u_{i}^{n}\right | ^4}{\left (\sum_{i} \left | u_{i}^{n}\right | ^2 \right )^2} .
\end{align}
In Fig.\ref{fig4}(a), we show the IPRs for all of the eigenstates.
It turns out that the localization property of the quantum walker is mainly determined by the IPRs of the edge states since all of the rest of eigenstates are delocalized. In Fig.\ref{fig4}(b), the IPR of one of the edge states is shown. It is found that for $\phi=0$ and $T=2$, the IPR of the edge state increases as the off-diagonal modulation grows stronger.

For comparison, we also show in Fig.\ref{fig1}(d-f) the dynamics of the quantum walker in a commensurate off-diagonal AAH lattice with $\phi=0.6\pi$ where the model is topological trivial with $\gamma=0$ and thus there is no edge state in the system's spectrum. It is found that the quantum walker is well delocalized as the strength of the off-diagonal modulation grows from $\lambda_{od}=0.1$ to $\lambda_{od}=0.9$. The variation of the off-diagonal modulation only slightly affects the expansion speed of the quantum walker. In Fig.\ref{fig1}(g-i), the dynamics of the quantum walker in a commensurate off-diagonal AAH lattice under periodic boundary condition is shown. The quantum walker shows no localization phenomenon since no edge state exists in the off-diagonal AAH lattice with periodic boundary condition even for the phase of $\phi=0$.

\begin{figure}[t]
\includegraphics[width=0.5\textwidth]{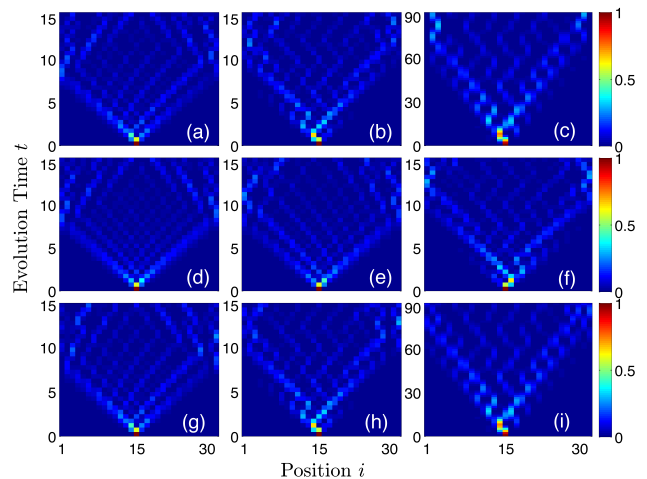} \caption{(Color online)
Single-particle quantum walks on a one-dimensional period-2 off-diagonal AAH lattice with $L=30$.
The quantum walker is initially positioned on the center site $15$.
(a-c) $\phi=0$, open boundary condition.
(d-f) $\phi=0.6\pi$, open boundary condition.
(g-i) $\phi=0$, periodic boundary condition.
The first column is corresponding to $\lambda_{od}=0.1$, the second is  $\lambda_{od}=0.3$ and the third is $\lambda_{od}=0.9$.
}
\label{fig5}
\end{figure}

\begin{figure}[b]
\includegraphics[width=0.5\textwidth]{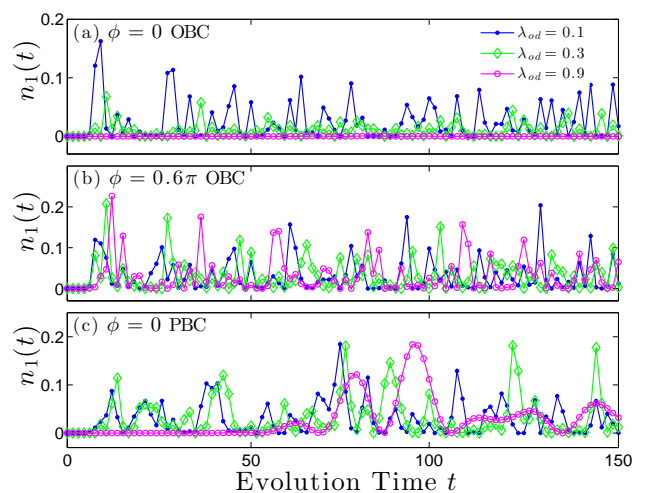} \caption{(Color online)
The time-dependent distribution of the quantum walker on the left boundary site for a relatively long time.
The quantum walker is initially positioned on the center site $15$.
(a) $\phi=0$, open boundary condition.
(b) $\phi=0.6\pi$, open boundary condition.
(c) $\phi=0$, periodic boundary condition.}
\label{fig6}
\end{figure}

Secondly, we investigate the dynamics of the quantum walker initially located on the lattice sites inside the bulk. As is shown in Fig.\ref{fig5}(a-c), for open boundary condition and phase
$\phi=0$, the quantum walker initiated from the center site expands ballistically and no localization phenomenon is shown as the strength of the off-diagonal modulation grows from
$\lambda_{od}=0.1$ to  $\lambda_{od}=0.9$. However, close and careful observation reveals an intriguing effect of the exotic edge state. If we focus on the two boundary sites of the lattice in Fig.\ref{fig5}(a-c), we will find that as the strength of the off-diagonal modulation increases, the distribution of the quantum walker on the boundary sites decreases gradually and finally disappears.
This may be seen as a \emph{repulsion} effect of the edge states and its strength is determined by the localization properties of the edge states. To be more clearly, we show the time-dependent distribution $n_{1}(t)$ of the quantum walker on left edge of the lattice for a long time period.
It is evident that for strong off-diagonal modulation $\lambda_{od}=0.9$, the quantum walker is repelled from reaching the boundary site as the distribution on site $1$ remains zero all the time, see Fig.\ref{fig6}(a).
Here in Fig.\ref{fig5} and Fig.\ref{fig6}, the lattice size is chosen to be $L=30$ for clarity.

Conversely, for open boundary condition with phase $\phi=0.6\pi$ and periodic boundary condition with $\phi=0$ when there is no edge state in the spectrum of the off-diagonal AAH model,  the quantum walker could reach the boundary sites easily, see Fig.\ref{fig5}(d-i) and Fig.\ref{fig6}(b-c).
The increasing of the off-diagonal modulation only affects the expansion speed of the quantum walker.

In a word, we have shown that the existence of exotic edge states in a period-2 off-diagonal AAH model have an interesting \emph{trapping} effect on the quantum walker initiated from the boundary sites of the lattice making the quantum walker localized and also, an intriguing \emph{repulsion} effect on the quantum walker set out from lattice sites inside the bulk prohibiting the quantum walker from reaching the boundary sites.
These two interesting effects should be observable with existing experimental  platforms, for example, an array of coupled photonic waveguides written in bulk glass using femtosecond laser microfabrication technology as used in \cite{Kraus, Lahini09}.
And they may have potential applications in the designing of micro-architectures for quantum information and quantum computing.  Imagine that two optical signals, one is injected into the phontonic waveguide at the boundary, the other is injected into a phontonic waveguide in bulk. By modulating $\lambda_{od}$ or phase $\phi$, these two signals could be made to meet each other or transmit separately.

\begin{figure}[b]
\includegraphics[width=0.48\textwidth]{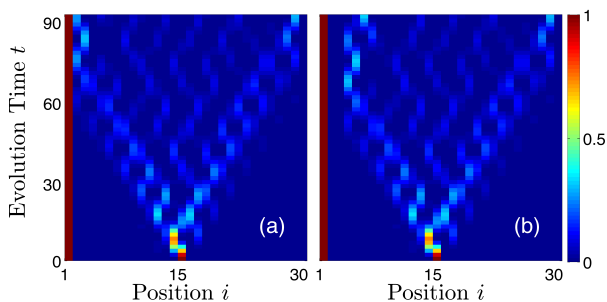} \caption{(Color online)
Quantum walks of two identical fermionic particles with nearest-neighbor interaction on a one-dimensional period-2 off-diagonal AAH lattice with $L=30$ under open boundary condition.  The phase $\phi=0$ and $\lambda_{od}=0.9$. One quantum walker is initially positioned on the left boundary site, and the other is located on the center site $15$.
(a) $V=0$.
(b) $V=1$.
}
\label{fig7}
\end{figure}

\section{Two-particle quantum walks}

\begin{figure}[t]
\includegraphics[width=0.46\textwidth]{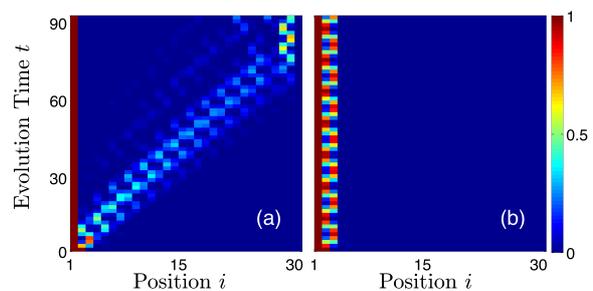} \caption{(Color online)
Quantum walks of two identical fermionic particles with nearest-neighbor interaction on a one-dimensional period-2 off-diagonal AAH lattice with $L=30$ under open boundary condition.  The phase $\phi=0$ and $\lambda_{od}=0.9$. The initial state is prepared on  $\left | 1,2 \right \rangle$.
(a) $V=0$.
(b) $V=1$.}
\label{fig8}
\end{figure}

In this section, we turn to investigate the continuous-time quantum walks of two identical fermionic particles on the commensurate off-diagonal AAH lattice with $T=2$.  As is shown in Eq.(\ref{H}),
the nearest-neighbor interaction between the two identical fermionic particles is considered.
We mainly focus on the effect of nearest-neighbor interaction on dynamics of the two quantum walkers setting out from different initial states.
Both in Fig.\ref{fig7} and Fig.\ref{fig8}, open boundary condition is adopted, the phase is set to $\phi=0$ and the strength of the off-diagonal modulation is set to be $\lambda_{od}=0.9$ when the \emph{trapping} and the \emph{repulsion} effects of the exotic edge states come into force on the dynamics of the quantum walkers. For clear visibility, the length of the AAH lattice is set to be $L=30$.

\begin{figure}[b]
\includegraphics[width=0.46\textwidth]{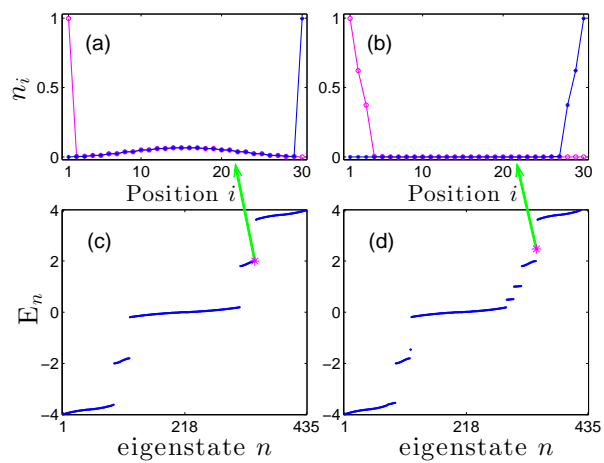} \caption{(Color online)
(a,b) Density distributions of eigenstates corresponding to the eigeneneries in (c,d) denoted by star symbols. 
(c,d)Eigenenergies in ascending order for period-2 off-diagonal AAH lattice with $\lambda_{od}=0.9$, $\phi=0$, and $L=30$ under open boundary condition. The nearest-neighbor interaction is
(c) $V=0$;
(d) $V=1$.}
\label{fig9}
\end{figure}

At first, we consider the case that
one quantum walker is initially located on an edge site of the AAH lattice and the other one is positioned on a site inside the bulk of the lattice. In Fig.\ref{fig7}, the initial state is chosen to be
$\left | 1,15 \right \rangle$. According to the discussions on single-particle quantum walks, we can infer that if no interaction is considered ($V=0$), the quantum walker on the edge site will be localized and the other quantum walker initiated from the center site will expand inside the bulk. They will propagate separately. This is exactly the picture shown in Fig.\ref{fig7}(a).
In Fig.\ref{fig7}(b), the nearest-neighbor interaction is set to $V=1$. It is found that the nearest-neighbor interaction dramatically enhances the \emph{repulsion} effect of the exotic edge states existing in the spectrum of the commensurate off-diagonal AAH model.
The region that can be reached by the quantum walker initiated from inside the bulk is evidently compressed.

Then we investigate the quantum walks of two identical fermionic particles initially located on the leftmost two lattice sites of the off-diagonal AAH lattice, i.e., the initial state is prepared as $\left | 1,2 \right \rangle$. Similarly as in Fig.\ref{fig7}(a), when the strength of the nearest-neighbor interaction $V$ is zero, the two quantum walkers transmit on the edge and inside the bulk respectively, see Fig.\ref{fig8}(a). However, in Fig.\ref{fig8}(b) we show that the quantum walker on the second site can be firmly \emph{pinned} by the quantum walker on the boundary site when the nearest-neighbor interaction is set to $V=1$. This is another interesting phenomenon that may have potential applications in microarchitecture designing.

These intriguing behaviors of quantum walkers are intimately related to the band structure and the eigenstates of the commensurate off-diagonal AAH model. In Fig.\ref{fig9}(c), we show the eigenenergies in ascending order for the off-diagonal AAH lattice with $\lambda_{od}=0.9$, $V=0$, $\phi=0$ and $L=30$ under open boundary condition. 
The density distributions of two typical eigenstates which contribute to the corresponding dynamical behavior of quantum walks in Fig.\ref{fig7}(a) and Fig.\ref{fig8}(a) are shown in Fig.\ref{fig9}(a).
The feature of these eigenstates is that half of the density of the quantum walker dwelts on the single boundary site and the other half of the density distributes among the rest of lattice sites.
The parameters in Fig.\ref{fig9}(d) is the same as in Fig.\ref{fig9}(c) except the nearest-neighbor interaction $V=1$. Compared to Fig.\ref{fig9}(c) with $V=0$,  eigenstates with density distributions like those shown Fig.\ref{fig9}(b) are singled out  by the nearest-neighbor interaction. A small energy gap appears, see Fig.\ref{fig9}(c). As shown in Fig.\ref{fig9}(b), almost all of the quantum walkers are distributed among a small region surrounding the lattice boundary. 
These eigenstates contribute to the exotic \emph{pinning} effect demonstrated in Fig.\ref{fig8}(b).

\section{CONCLUSIONS}

In summary, we have investigated the single-particle and two-particle continuous-time quantum walks on a one-dimensional commensurate off-diagonal AAH lattice.
Especially, the effect of the topological property of the commensurate off-diagonal AAH model on the dynamics of the quantum walks has been addressed. In the parameter region where the model is topologically nontrivial, edge states will emerge in the spectrum of off-diagonal AAH lattice under open boundary condition. The quantum walker initiated from the boundary site of the AAH lattice will be localized when the IPRs of the edge states are large, which can be modulated by the strength of the off-diagonal modulation. When the quantum walker initially set out from a lattice site inside the bulk, it will encounter an intriguing \emph{repulsion} effect of the exotic edge states. For quantum walks of two identical fermions, it is found that the nearest-neighbor interaction
could dramatically enhance the \emph{repulsion} effect of these exotic edge states.
Also, an interesting \emph{pinning} effect is revealed in the quantum walks of two identical fermions initially positioned on the two leftmost sites.
These effects may be observed experimentally in one-dimensional array of photonics waveguides\cite{Kraus, Lahini09}, double-well potentials\cite{sebby,foiling, wirth}, optical lattices\cite{billy,roati}  or semiconductor structures\cite{sarma,merlin}.  And they may have prosperous applications in the designing of microarchitectures for quantum information and quantum computing.

\begin{acknowledgments}
This work is supported by NSF of China under Grant Nos. 11474189, 11404199 and 11674201,  NSF for youths of Shanxi Province No. 2015021012, and research initiation funds from SXU No. 216533801001. 
S.C. is supported by NSF under Grant Nos. 11425419 and 11374354.
The work is partially completed during the visit in ANU which is sponsored by CSC(No. 201508140015). L. W. would like to thank the hospitality of Profs. V. Bazhanov, A. Truscott and X.-W. Guan during the stay in ANU.
\end{acknowledgments}


\end{document}